\documentstyle[12pt]{article}
\vspace{5cm}
\title{Description of relativistic heavy--light quark--antiquark systems via
Dirac equation} \vspace{2cm} \author{V.D. Mur, V.S. Popov, Yu.A.Simonov, and
V.P.Yurov} 
\begin{document}
\date{} \maketitle
\centerline{Institute of Theoretial and Experimental Physics}
\centerline{117259, Moscow, Russia}
\centerline{Submitted 25 June 1993}
\vspace{1cm} \begin{abstract}
Starting from the QCD Lagrangian and taking into account both perturbative and
nonperturbative effects, we use the  method of vacuum correlators to derive the
Dirac equation (rigorously for the Coulomb interaction and heuristically for
the
confining potential) for a system consisting of a light quark and heavy
antiquark.  As a result the confining potential is a Lorentz scalar, and the
Coulomb part the fourth component of a 4-vector. The energy spectrum of the
Dirac equation is considered for these
potentials. Numerical calculations of energy eigenvalues $E=E_{n\kappa}$
are  performed, and  some exact solutions of the Dirac equation in the
case $E=0$ are found.
An effective-potential  method convenient for
qualitative study of the solutions of the Dirac equation is developed.
The connection with experimental spectra of $D$- and $B$-mesons is
briefly discussed.

\end{abstract}
\def\thefootnote{\arabic{footnote} )}

\section{Introduction}

\hspace{.5cm}
The quark--antiquark systems consisting of one heavy quark (antiquark)
$Q$ and one light antiquark (quark) $\bar{q}$ are  QCD analogs of the
hydrogen atom and thus are of fundamental importance.

Recently the issue of the heavy--light bosons has become a topic
of  vivid interest in both  analytic and Monte Carlo QCD studies [1].

{}From the theoretical viewpoint the interest in  heavy--light systems stems
from  several considerations. First, in the limit of one infinitely heavy
quark, one hopes to get the dynamics of a light quark in the external
field of a heavy one.  That would be similar to the picture of the
hydrogen atom.

Second, since the external field is time-independent, one may hope to
obtain a static potential in QCD together with spin-dependent forces, as
has been  done for heavy quarkonia [2].

An important issue in this connection is the Lorentz nature of the confining
part
of the potential. Arguments in favor of a scalar nature are found in
the form of inequalities [3] and also in the form of spin-dependent terms
[2,4].

Third, in the heavy--light system one may study how the chiral symmetry
breaking
(CSB) affects the spectrum. When one quark mass is vanishing, in the chiral
symmetric
case the spectrum would consist of parity doublets, and the CSB would lift
the degeneracy.

Fourth, using the Dirac equation we implement explicitly relativistic
dynamics and can study relativistic properties of the spectrum, e.g., in the
case of a vanishing quark mass, and also relativistic spin splittings in the
spectrum.

In particular, in our case the spin symmetry now being widely discussed
[1] is present, since the spin of the heavy quark is decoupled. Hence every
state of the Dirac equation with a given $j$ and parity corresponds to  two
almost degenerate states of the $(q\bar{Q})$ system with ${\bf j}$ and the
heavy quark spin $S = 1/2$ adding to $J=j\pm1/2$.

Our final point is that the Dirac equation description yields a dynamical
framework
for the heavy--light mesons which can be used to calculate meson matrix
elements and
 form factors to compare with experiment and semiphenomenological approaches
now widely used  in this context [1,5].

Here we attempt to derive explicitly the Dirac equation for the heavy--light
quark system starting from the QCD Lagrangian and incorporating both
perturbative and nonperturbative effects in the framework of the vacuum
correlator method (VCM) [6--8].

In the course  of the derivation we use some approximations, which are
clearly stated at each step, e.g., neglect of virtual quark pairs
(the quenching approximation), keeping the lowest (quadratic) cumulant in
the cluster expansion, and the "limit of local dynamics" [2], $T_g
\rightarrow 0$, where $T_g$ is  the gluonic vacuum correlation length. We
discuss physical implications of these approximations in the main text
below.

Even with those approximations we are unable to prove rigorously the
appearance of a static scalar confining potential, but we give
strong arguments in favor of it. The situation is much better for the
perturbative
part, and the existence of a external vector potential in the limit of one
heavy quark mass is explicitly shown.

At this point we start with the Dirac equation containing a vector Coulomb
part and a scalar confining part, and we study properties of its spectrum.
For comparison we  also consider two other cases:  i)the equation has a
vector confining part and a Coulomb part; ii) both parts of the potential
transform as scalars. We are explicitly interested in the limit of a
vanishing light quark mass and the spin dependence of the energy
eigenvalues.

We show explicitly that a reasonable spectrum occurs only for a scalar
confining part. In this case the scalar potential breaks explicitly
chiral symmetry, and  states with opposite parities are not
degenerate.
In the case of a vector confining part, only quasistationary states are
found.

We write possible general symmetry properties of spectral
states and also the corresponding properties in the particular cases of  a
zero scalar potential, a zero vector potential, and a zero mass.

We also find some exact analytic solutions of the Dirac equation which
occur for $E = 0$.
This enables us to find an equation, which determines that value of the
Coulomb constant $( \zeta = \zeta_{nx} )$, for which the $(njl)$ discrete
energy level reaches the value $E = 0$. We note an analogy of this problem
with the relativistic Coulomb problem of an electron in the field of a
heavy supercritical nucleus with $Z > 137$ [9--11].

The paper is organized as follow.
In Sec. 2 we use the Feynman--Schwinger representation of the
quark--antiquark Green function to derive the limit of one infinit mass.

In Sec. 3 we compare this limit with the Dirac equation and discuss
the Lorentz nature of the confining interaction.

In Sec. 4 we discuss the properties of the spectrum of the Dirac
equation.

In Sec. 5 we discuss symmetry properties of the spectrum, in particular
chiral properties.

In Sec. 6 we discuss numerical results for the spectrum of the Dirac
equation in the limit of a vanishing small quark mass.

In Sec. 7 we investigate the exact analytic solutions of the Dirac
equation which occur for $E = 0$.

In Sec. 8 we summarize our results and compare the Dirac spectrum
obtained here with experiment and other approaches. This enables
us to also discuss large-mass corrections.

\section {Feynman--Schwinger representation for the
heavy--light  Gre\-en's function}

\hspace{.5cm}
Consider the quark--antiquark system of one heavy antiquark (with mass
$m_2$) and a light quark of much smaller mass $m_1$, which we neglect in
some cases below.

The Green's function $G$ of such a system, with the quarks initially at
points $y$, $\bar{y}$ and finally at points $x$, $\bar{x}$, is given by the
path integral over fermionic and gluonic fields A in ref. [8].
 Evaluating the fermionic integrals and neglecting the
fermionic determinants (the quenching approximation) for simplicity, one
arrives at the amplitude depicted graphically in Fig.1, where $\Delta_i$ is
the quark Green's function with quark mass $m_i$,  and the $\Gamma$'s
belong to the initial and final wave functions:

$$\Psi(y, \bar{y}) = \bar{q}(y) \Gamma(y, \bar{y}) q(\bar{y}),$$
 with a similar expression for $\Psi(x,\bar{x})$. Thus we can
write \begin{equation} G(x\bar{x}|y\bar{y}) =
\langle \mbox{\rm tr}[\Gamma(\bar{x},x)\Delta_1(x,y)\Gamma
(y,\bar{y})\Delta_2(\bar{y},\bar{x})]\rangle .
\end{equation}
The angular brackets here denote the integration over gluonic fields.
This integration we perform using the cluster expansion [6,7].
We  arrive at the following
expression involving proper-time and path integrals for the quark
$(s,z)$ and the antiquark $(\bar{s},\bar{z})$:
\begin{eqnarray}\nonumber
G(x\bar{x}|y\bar{y}) = \int\limits^{\infty}_0 ds \int\limits^{\infty}_0
d\bar{s} Dz
D\bar{z} e^{-K-\bar{K}} \times\\
%\nonumber
 \times\mbox {\rm tr} [\Gamma_x(m_1 - \hat{D}^{(1)})
\gamma_{z\bar{z}} \Gamma_y (m_2 - \hat{D}^{(2)})],
\end{eqnarray}
where $$K = m^{2}_{1}s + \frac{1}{4} \int\limits^{s}_{o} \dot{z}^2 d \tau,$$

$$\bar{K} = m^{2}_{2}\bar{s} + \frac{1}{4} \int\limits^{\bar{s}}_o
\dot{\bar{z}}^2 d
\bar{\tau}.$$
In (2), $\Gamma_x$, $\Gamma_y$ refer to the Lorentz structures in the
initial and final states, and $D^{(i)}$
and $\gamma_{z\bar{z}}$ are given in ref. [7].  We rewrite the latter here in
the following form (in the lowest order of a cluster expansion, where
perturbative and nonperturbative contributions factorize [8]):
\begin{equation}
\gamma_{z\bar{z}} = \exp \left[ -\frac{g^2}{2}\int df_{\mu\nu}(u)
df_{\rho\lambda} (u') \langle F_{\mu\nu}(u)F_{\rho\lambda}(u') \rangle \right],
 \end{equation} where
\begin{equation} d f_{\mu\nu} (u) \equiv d s_{\mu\nu}(u) -
i\sigma^{(1)}_{\mu\nu} d \tau + i \sigma^{(2)}_{\mu\nu} d \bar{\tau} ,
\end{equation} and $\langle F$ $F\rangle $ is the gluonic correlator. The
latter
can be split into perturbative and nonperturbative parts as in ref. [7].

      Now we discuss the limit of one heavy mass, $m_2 \rightarrow \infty$.
In this case, when $m_2$ is much larger than the interaction strength,
 particle 2 is moving along the straight-line trajectory
%\end{document}

\begin{eqnarray}
\bar{z}_{\mu} = \bar{y}_{\mu} + \frac{\bar{x}_{\mu}-\bar{y}_{\mu}}
{\bar{s}}\tau, \;\;\;   0 \leq \bar{\tau} \leq \bar{s} ,\nonumber \\
\bar{x}_4 - \bar{y}_4 \equiv T,~~  \frac{T}{\bar{s}} = 2m_2,~~
d\bar{\tau} = \frac{d\bar{z}_4} {2m_2};
\end{eqnarray}
   $\bar{x}_i = \bar{y}_i = R_i$ is the position in space of particle 2.

One can see that in this limit the spin interaction of particle 2 (terms
$\sigma^{(2)}d\bar{\tau}$) is  $\sim  O(1/m_2)$  and can be
neglected.

The perturbative part of the cumulant $\langle \langle F(x)F(y)\rangle \rangle$
can be written
as [6,7]
%\begin{equation}
 $$g^2\int d\sigma_{\mu\nu}(u) \int
d\sigma_{\rho\lambda}(u') \langle \langle F_{\mu\nu}(u) F_{\rho\lambda}(u')
\rangle \rangle_{pert}
=$$
%\begin{equation}
 $$= g^2 \int\limits_{C} dz_{\mu} \int\limits_{C} dz'_{\rho} \langle
A_{\mu}(z)A_{\rho} (z') \rangle _{pert} =$$
\begin{equation}
  =\frac{C_{2}g^{2}}
{4\pi^2} \int\limits_C dz_{\mu} \int\limits_C \frac {dz'_{\mu}} {(z-z')^2},
\end{equation}
where $C_2$ is the quadratic Casimir operator, $C_2(N_C=3) = 4/3
$, and $C$  is the closed contour depicted in Fig.1. The integral
in (6) is divergent and is regularized by inserting a $Z$
factor in front of the exponential in (3) and introducing the
minimal distance $\delta$ of the points $\bar{z}$ and $\bar{z}'$ in (6)
(for details see [12]).

Therefore we can now study the situation when only the perturbative
interaction is present, and the Green's function looks like
\begin{eqnarray}\nonumber
G(x\bar{x}|y\bar{y}) = \int\limits^{\infty}_0 dsDz \exp\left(-m^{2}_{1} s -
\frac{1}{4} \int\limits^{s}_{0} \dot{z}^2 d \tau\right) \mbox {\rm tr}
[\Gamma_x(m_1 +
\frac {1}{2}\dot{z}) \times \\
\nonumber \times Z \exp \left[-g^2 \int(ds_{\mu\nu}(u) + \frac {1}{i}
\sigma^{(1)}_{\mu\nu}d\tau) ds_{\rho\lambda} (u') \langle \langle F_{\mu\nu}(u)
F_{\rho\lambda}(u') \rangle \rangle_{pert} \right] \times
\\
%\nonumber
\times \Gamma_y (m_2(1-\gamma_4))],
\end{eqnarray}
where we have neglected terms representing the  self-interaction of particle 1.

Our goal now is to rewrite the exponential function in (7) in the form

\begin{equation}
\exp \left[ig \int(\bar{A}_{\mu}(z)dz_{\mu} + \frac{1}{i} \sigma^{(1)}_{\mu\nu}
\bar{F}_{\mu\nu}  d\tau )\right]
\end{equation}
(a similar derivation for QED was done in [13]),where we  have
defined (following [6])
%\begin{equation}
$$\bar{A}_{\mu}(z) = ig \int \langle A_{\mu}(z)A_{\rho}(z') \rangle dz_{\rho}'
=
\frac{ig}{4\pi^2} \int \frac{dz'_{\mu}}{(z-z')^2} =$$
\begin{equation}
= \frac{ig}{4\pi}\frac{\delta_{\mu 4}}{\mid{\bf z} - {\bf R}\mid} C_2.
\end{equation}

Similarly,
\begin{equation}
\bar{F}_{\mu\nu} = \partial_{\mu}\bar{A}_{\nu} - \partial_{\nu}
\bar{A}_{\mu}.
\end{equation}
One can also verify that the explicit
expression for $\langle \langle F$ $F \rangle \rangle_{pert}$, which one
obtains from the Feynman
gauge propagator [6], leads to
\begin{equation} \langle A_{\mu}(z)A_{\nu}(z') \rangle _{pert} =
\frac {1} {4\pi^2} \frac {\delta_{\mu\nu}} {(z-z')^2}. \end{equation} Hence
we can write the $q\bar{q}$ Green's function, keeping only the perturbative
interaction:
\begin{equation}
G(x\bar{x}|y\bar{y}) = \mbox {\rm tr} [\Gamma_x G(x,y) \Gamma_y
m_2(1-\gamma_4)],
\end{equation}
where we have defined
%\begin{equation}
$$G(x,y) = \int dsDz[m_1-\hat{D}(\bar{A})]\times $$
$$\times\exp\left(-m^{2}_{1}s-\frac{1}{4}\int\limits ^{s}_{0}\dot{z}^2d\tau +
ig\int\limits^{x}_{y}\bar{A}_{\mu}dz_{\mu} + g\int\limits^{s}_{0}
\sigma_{\mu\nu}\bar{F}_{\mu\nu}d\tau \right)  =$$
 $$=\langle x \mid\int\limits^{\infty}_{0} ds [m_1 - \hat{D}(\bar{A})]
\exp\left\{-s[m^2-\hat{D}^2(\bar{A})]\right\} \mid y \rangle =$$
\begin{equation} =\langle x \mid (m_1 + \hat{D}(\bar{A}))^{-1} \mid y \rangle .
\end{equation} We observe that $G(x,y)$ satisfies the Dirac equation
\begin{equation} [m_1 + \hat{D}(\bar{A})] G (x,y) = \delta^{(4)}(x-y).
\end{equation}

In the next section we  study the effect of the nonperturbative
interaction.

\section{Confining force and the Dirac equation}

\hspace{.5cm}
We now turn to the nonperturbative interaction, which provides
confinement due to the presence of the special (Kronecker) structure
in the quadratic cumulant [6]. Again neglecting the self-interaction
of light particle, we find an equation of the same form  as
(7), but we should add  $\langle \langle F$ $F \rangle \rangle_{pert}$ also the
nonperturbative part $\langle \langle F F \rangle \rangle_{nonpert}$, which
can be written as [6]
\begin{eqnarray}\nonumber
\langle \langle F_{\mu\nu}(u) F_{\rho\lambda}(u') \rangle \rangle =
(\delta_{\mu\rho}
\delta_{\nu\lambda} - \delta_{\mu\lambda} \delta_{\nu\rho}) D
(u-u') +
\\
%\nonumber
+ \frac{1}{2}\left\{\frac{\partial}{\partial u_{\mu}}
[(u-u')_{\rho} \delta_{\nu\lambda} - (u-u')_{\lambda}  \delta_
{\nu\rho}] + \mu\rho \leftrightarrow \nu\lambda \right\} D_1(u-u').
\end{eqnarray}

It has been shown [6,7] that only $D$ (and not $D_1$) yields confinement
(the area law of the Wilson loop). We concentrate first on this
term, disregarding also the spin term $\sigma^{(1)}_{\mu\nu}$ in (7).
We have
\begin{equation}
\int ds_{\mu\nu}(u) \int ds_{\rho\lambda}(u') \langle \langle F_{\mu\nu}(u)
F_{\rho\lambda}(u') \rangle \rangle \rightarrow \int ds_{\mu\nu}(u) ds_
{\mu\nu}(u')D(u-u') + ... \; .
\end{equation}
Using the same arguments as in [6], one can see that we get the
area-law exponential function in (7). Namely, for large area we have
\begin{equation}
\exp (-\sigma S_{min}),  ~~~~    \sigma = \frac{1}{2}\int\!\int d^2 u D(u),
\end{equation}
where $S_{min}$ is the minimal area inside the contour formed by the
straight-line trajectory of the heavy particle and the path $z(\tau)$
of the light one.

We are now facing some fundamental questions: i) Can the term (17)
 be associated with a local potential $V$, acting on particle 1?
ii) What are the Lorentz properties of this potential --- does it
transform as a scalar or as a vector?

To answer  question i) we  follow the arguments  given in [2].
We must therefore return to the exponential function in (3), defining the
dynamics of the system. The integral in the exponential function in (3) is
over the surface inside the quark and antiquark trajectories; the
characteristic length of these trajectories is $T_q$, a period of quark
orbiting.

Being at some point $z(\tau_0)$ on the trajectory, a quark is influenced
by the fields and through them by its partner. The radius of nonlocality
of the fields is given by the correlation length $T_g$, defining
behavior of $\langle F(u)F(u')\rangle $, i.e., by the functions
$D((u-u')/T_g)$, $D_1((u-u')/T_g)$.

Therefore the criterion of local dynamics is $T_q \gg T_g$  [2,14].

In the opposite case, $T_q \leq T_g$, the quark "feels" all the fields
and also the motion of the antiquark during all its history. This is the
nonlocal dynamics, which can be treated by the $QCD$ sum rules [15].
Lattice calculations yield $T_g \sim 0.2 \div 0.3$  fm [16],
while $T_q$ for both the light and heavy quarks is $T_q \geq 1$  fm.
Therefore the actual situation is close to the local dynamics.

Regarding the first point, we  proceed as in [8, 17], forming
the minimal surface via the connection of $\gamma \equiv
\bar{\tau}/\bar{s} = \tau/s =t/T$, where $t =\bar{z}_4 - \bar{y}_4$.

In this case, combining all exponential functions in (13) and (17), we obtain
\begin{equation}
B=\int\limits^{T}_{0} dt \left[\frac{m^2_1}{2\mu_1} + \frac{\mu_1}{2} +
\frac{\mu_1}{2}\dot{r}^2_{\alpha}(t) + \sigma \int\limits^{1}_{0} d\beta
\sqrt{\dot{w^2}r^2-(\dot{w_{\mu}}r_{\mu})^2}\right] ,
\end{equation}
where we have defined, as in [8,18],
\begin{eqnarray}      \nonumber
w_{\mu}(t)=z_{\mu}(t)\beta + \bar{z}_{\mu}(t)(1-\beta),~~~
\dot{w}_{\mu}= \frac{\partial}{\partial t}w_{\mu},\\
%\nonumber
r_{\mu}(t)= z_{\mu}(t)-\bar{z}_{\mu}(t),~~~ 2\mu_1=T/s.
\end{eqnarray}

Taking (5) into account, we have
\begin{eqnarray}
\nonumber
w_i(t)=z_i(t) \beta+R_i(1-\beta),~~ R_i=\mbox{\rm const} , \\
%\nonumber
w_4(t)=z_4(t) \beta+(\bar{y}_4+t)(1-\beta) , \\
\nonumber
\dot{w}_i(t)=\dot{z}_i(t) \beta= \dot{r}_i(t) \beta,~~~
\dot{w}_4=\dot{z}_4(t) \beta+ 1- \beta .
\end{eqnarray}

Following the same procedure as in [8,18], we
obtain from $B$ (considered as an action) the proper-time Hamiltonian,
which depends on the dynamical mass $\mu_1$:
\begin{equation}
H(\mu_1)= \frac{m_1^2}{2\mu_1}+\frac{\mu_1}{2}+ \sigma \mid{\bf r}\mid +
\frac{1}{2\mu_1}\left(-\frac{\partial^2}{\partial{\bf r}^2}\right) .
\end{equation}

The eigenvalue of $H(\mu_1)$ is to be minimized with respect to
$\mu_1$; the final value of $\mu_1$ is determined in this manner.

Thus we see that  the nonperturbative correlator $\langle \langle FF
\rangle \rangle_{nonpert}$ does indeed
yield a potential-type term in the proper-time Hamiltonian, provided
$T_g$ is small [2] and the expansion of the square-root term in $\beta$
is done as in [18] (the corresponding error is  $\sim$10\% ).

It is more difficult to answer the other question --- about the  Lorentz
nature of this potential --- since Eq. (21) is written in the c.m. system, and
we do not know how this expression transforms under Lorentz boosts. To get a
partial understanding of this point, let us compare how the vector and
 scalar potentials enter  the Feynman--Schwinger representation for the
 Green's function of a scalar quark [i.e., neglecting a spin term
 $\sim\sigma^{(1)}$ in (13)].

 For the scalar case we can write

\begin{equation}
G(x,y)\sim \int ds Dz \exp \left[-\int\limits^{s}_{0} V_s(z(\tau)) d\tau- m^2s-
\int\limits^{s}_{0}\frac{\dot{z}^2}{4}d\tau\right] ,
\end{equation}
while for the vector case we have
\begin{equation}
G(x,y)\sim \int ds Dz \exp
\left(-m^2s-\int\limits^{s}_{0}\frac{\dot{z}^2}{4}d\tau + i
\int\limits{s}_{0}V_{\mu} d z_{\mu}\right) .
\end{equation}

The main difference lies in the fact that trajectories of the
$Z$-type depicted in Fig.2 (virtual pair creation) give different
contributions in the scalar and vector case. For the scalar case the proper
time intervals $\Delta \tau_1$, $\Delta \tau_2$, and $\Delta\tau_3$ are
positive and add together, thereby suppressing  the pair creation, while
in the vector case $\Delta z^{(1)}_{4}, \Delta z^{(2)}_{4}$, and $\Delta
z^{(3)}_{4}$ strongly cancel each other, making pair creation easy. In other
words, for the scalar potential all parts of a trajectory add
arithmetically, while for the vector potential should take a vector sum of
all intervals along the trajectory.

Looking at our prototype of the potential, the last term in (18), one can see
that in this case one actually adds a scalar quantity $\dot{\sigma}dt
\sqrt{\ }$ (we recall that $dt$ is  not the fourth component of a vector, but
rather the proper time $d\tau$, which is scalar).
Thus we seem to be justified in treating the confining force as a scalar and
not
a vector.

There are additional arguments in favor of this conclusion.

First of all, in the nonrelativistic  derivation of spin-dependent forces in
nonperturbative QCD [2], the sign and magnitude of the spin--orbit term
depend on whether the confining potential is chosen as a scalar or a
vector [4].  In  Appendix B we reproduce the spin-dependent terms obtained
in [2], for the  case of two nonrelativistic quarks.
 One can notice that the asymptotically dominant contribution comes
from $V'_1$, which contains only a scalar contribution and yields the
negative coefficient of the spin--orbit term ${\bf L}{\bf S}$, characteristic
of Thomas precession.

Results obtained in [2] through the use of the cluster expansion and the
area law unambigiously predict the spin--orbit force corresponding to the
Thomas precession term, which was also introduce a previously using the
string picture [19]. All that corresponds to  scalar confinement.

Second, there are independent arguments in [3] which lead  to
 inequalities  which are satisfied for the scalar confining potential, and
not satisfied for vector case.  Finally, the vector confining potential is
believed to cause the Klein paradox [10,11] . In Sec. 4 and 6 we study
both scalar and vector confining potentials and show analytically
and numerically that, indeed, only in the first case does one obtain a
physically
consistent spectrum, corresponding to confinement while in the second case
one has only quasistationary states. This situation  is related to the
Klein paradox [10,11].  Thus we give strong arguments in favor
of the scalar confining potential.  Unfortunately at the moment we are still
unable to derive from the  Feynman--Schwinger representation the Dirac
equation with the scalar potential, in the way we have done  for
a perturbative Coulomb interaction.

Therefore in the next sections we simply postulate the Dirac equation
with a confining potential of the scalar type (or vector type --- to
 check  its inconsistency).

\section{Properties of the spectrum of the Dirac equation}

\hspace{.5cm}
As we  showed in the previous sections, the color Coulomb interaction between
quarks is of a vector nature. We  have argued that the confining
interaction is a scalar.
It is instructive, nevertheless, to consider both possibilities for the
interactions, scalar $U$ and vector $V$, and to study the properties of
solutions of the Dirac equation in these cases.

 Assuming spherical symmetry, we look for solutions of the Dirac equation in
the
form
\begin{equation}
\Psi = r^{-1}\left(\begin{array}{c}
     G(r)\Omega_{jl_M}\\
     iF(r)\Omega_{jl'_M}
\end{array}
\right) ,
\end{equation}
where $l+l'=2j$, and the $\Omega $'s are spherical spinors [20].

Equations for radial wave functions are
\begin{eqnarray}
\nonumber
\frac{dG}{dr} + \frac{\kappa}{r}G -
[E+m+U(r)-V(r)]F=0,\\
\frac{dF}{dr} - \frac{\kappa}{r}F +
[E-m-U(r)-V(r)]G=0 ,
\end{eqnarray}
where $E$ and $m$ are the energy and mass of the light particle ($m\equiv m_1$
in the notation of Sec. 2), and

\begin{equation}
\kappa = \left\{ \begin{array}{ll}
-(l+1),~~ & \mbox{if}~~~ j=l+1/2 , \\
     l,~~ & \mbox{if}~~~ j=l-1/2 ,
\end{array} \right.
\end{equation}
so that $|\kappa | =j+1/2= 1,2,...\; .$ We consider three choices.

a) Assume
\begin{equation}
U(r)=M^2r,~~ V(r)=-\frac{\zeta}{r},~~ M^2\equiv \sigma; ~~
\zeta=\frac{4}{3}\alpha_s.
\end{equation}

Putting $m=0$ and introducing the dimensionless variables $x\equiv Mr$ and
$\varepsilon =E/M$, we arrive at
\begin{eqnarray}
G'+ \frac {\kappa}{x}G-\left(\varepsilon+x+\frac {\zeta}{x}\right) F=0,\\
\nonumber
F'- \frac {\kappa}{x}F+\left(\varepsilon-x+\frac {\zeta}{x}\right) G=0
\end{eqnarray}
(hereafter the prime denotes the derivative with respect to $x$).

Let us find the asymptotics of the radial functions at zero and infinity. For
$x\rightarrow 0$, inserting the expansions $$G=Ax^{\nu}+...,~~~
F=Bx^{\nu}+...$$ into (28), we obtain
\begin{equation}
\nu =(\kappa^2-\zeta^2)^{1/2},~~
A/B=\kappa^{-1}(\kappa^2-\zeta^2)^{1/2}-1. \end{equation}

The wave function is regular at zero for $\zeta<\mid \kappa\mid
=j+1/2$ while for $\zeta>j+1/2$ there occurs a
 "collapse to the center" ( well known in quantum mechanics ) [21,22]. The
difficulty is actually of a formal character and can be removed
by introducing a cutoff of the Coulomb potential at small distances,
$r<r_0$ --- in the same way as in QED with the charge $Z>\alpha^{-1} = 137$
(see refs. [9--11, 23,24]).

The behavior of the wave functions at infinity is more complicated.
{}From (28) it follows that $G,F\sim \exp (\pm x^2/2)$ as
$x\rightarrow \infty$ , and the solution with a positive exponent is clearly
not admissible.

We accordingly write $G,F$ as
\begin{eqnarray} \nonumber
G\approx A' \exp [-(ax^2+bx)]x^{\beta}\left(1-\frac{c_1}{x}+
\frac{c_2}{x^2}+...\right),\\
%\nonumber
F\approx B' \exp [-(ax^2+bx)]x^{\beta}\left(1-\frac{c'_1}{x}+
\frac{c'_2}{x^2}+...\right).
\end{eqnarray}

Inserting these expansions into (28) we find, after some elementary but
cumbersome  calculations,
\begin{eqnarray}\nonumber
 A'=-B',~~ a=1/2,~~ b=0,~~ \beta=\varepsilon^2/2,~~\\
%\nonumber
c_1=(\zeta-1/2)\varepsilon,~~
c'_1=(\zeta+1/2)\varepsilon,... \; .
\end{eqnarray}
The following coefficients, $c_2, c_2', ... $ , can be calculated by means
of recurrence relations\footnote{Thanks are due to D.Popov for the
derivation of the relations and for verification of Eqs.$\;$ (29)--(31).}.

Thus the wave functions fall off at infinity mostly in the same way as in the
case of the harmonic oscillator in nonrelativistic quantum mechanics. The
solution satisfying the conditions obtained above, (29) and (30), exists
only for certain values of $\varepsilon=\varepsilon_n(\zeta, \kappa)$,
and the wave function is normalizable.  Therefore, for the Dirac equation only
a discrete spectrum  exists.

 This conclusion holds for any scalar potential which grows at infinity,

\begin{eqnarray}
G(r), F(r) \sim \exp \left[-\int\limits^{r}_{0}U(r)dr'\right],~~ r\rightarrow
\infty .
\end{eqnarray}

Even for the logarithmic potential the wave functions decrease
asymptotically faster than exponentially:
$$G, F_{(r\rightarrow \infty)}
\sim \exp (-gr\ln r) \mbox~~~ {\rm for}~~ U(r)=g\ln r . $$

b) When both interactions are of a vector type, i.e.,
\begin{eqnarray}
U(r)=0,~~ V(r)=-\zeta r^{-1}+M^2r,
\end{eqnarray}
the system of radial equations is
\begin{eqnarray} \nonumber
G'+ \frac{\kappa}{x}G-\left(\varepsilon+\frac{\zeta}{x}- x\right) F=0,\\
%\nonumber
F'- \frac{\kappa}{x}F+\left(\varepsilon+\frac{\zeta}{x}-x\right)G=0.
\end{eqnarray}

In this case, at $x \rightarrow 0$ we have
$$\nu =(\kappa
^2-\zeta^2)^{1/2},~~~
A/B=\kappa/[\kappa+(\kappa^2-\zeta^2)^{1/2}], $$
while the parameters in Eq.$\;$(30) assume the values
$$A'=iB',~~ a=-i/2,~~ b=i\varepsilon,~~ \beta=-i\zeta . $$

Here we chose a solution satisfying the Sommerfeld radiation condition,
i.e., having an outgoing wave at $r\rightarrow \infty$. Thus for
interaction (33) the discrete spectrum is absent (see also curve {\it 2} in
Fig.3), and all solutions of the Dirac equation, if any, are
quasistationary.  In the special case of a square well one can obtain
an exact analytic equation for the spectrum [25].  The width of these
quasistationary states determines the probability of  spontaneous pair
creation in  potential (33). This phenomenon is closely connected with
the  Klein paradox [10,11].  Physically it means that the problem
is now of the many-particle type, but actually the Dirac equation is still
applicable.

c) For completeness, consider also the case of purely scalar interactions,
\begin{eqnarray}
U=-\zeta r^{-1}+ M^2r,~~ V=0.
\end{eqnarray}

In this case we have at $r \rightarrow 0$, instead of (29),
\begin{equation}
\nu =(\kappa^2+\zeta^2)^{1/2},~~ A/B=
[\kappa-(\kappa^2+\zeta^2)^{1/2}]/ \zeta .
\end{equation}

Thus the "collapse to the center" does not occurs for arbitrary large
values of the Coulomb parameter $\zeta$. At infinity we obtain a
behavior which is analogous to (30):$$G, F \sim \exp (-M^2r^2/2);$$
therefore the spectrum is discrete.

Hence the character of the energy spectrum depends crucially on the assumption
about the nature of the confining interaction $M^2r$ (scalar or vector).

The reason can be understood qualitatively by using the method of an
effective potential [24,26].

System (25) corresponds in the quasiclassical approximation
(neglecting spin--orbit and spin--spin forces) to the following relation
between energy and momentum
\begin{equation}
p^2(r) = [E - V(r)]^2 - [m + U(r)]^2 + \kappa^2r^{-2} \equiv 2m
(\tilde{E} - U_{eff}),
\end{equation}
 where $\tilde{E}$ and $U_{eff}$ are the
effective energy and potential in the nonrelativistic Schr\"{o}dinger
equation:
\begin{eqnarray} \tilde{E} = \frac{1}{2m}(E^2 - m^2) \mbox{,~~~}
U_{eff}=\frac{1}{2m}(U^2-V^2) + U+\frac{\kappa^2}{2mr^2} .
\end{eqnarray}

In the nonrelativistic limit $(E \approx m \;\;\mbox{and}\;\; U, \mid V\mid
\ll m)$ we have $$U_{eff} \approx U + V+(l+1/2)^2/2mr^2.$$ Therefore, all
three cases, a)--c) correspond to the tunnel potential $$U_{eff} (r) =
-\zeta/r+ \sigma r,$$ which is often used in QCD.

However, for large $U$ and $V$, the situation is drastically different
for our cases a)--c). This can be understood from Fig. 3, where the
qualitative behavior of $U_{eff}(r)$ is shown.
Using the effective-potential method, one can also perform a
quantitative study of the problem (for details see Appendix A).
Applying the WKB method and taking into account Eq.$\;$ (37), one can easily
determine the asymptotics  of the wave functions at infinity.

For example, in  case a) we get
\begin{equation}
G,F \sim \exp \left( - \int\limits_0^x|p| dx\right),~~ |p| =
U+m-\frac{\varepsilon^2}{2U} +\frac{\varepsilon V}{U} +... \; .  \end{equation}

If $U(x) = gx^{\alpha}, \alpha>0$, and $V(x) = -\zeta /x$,
then at $x \rightarrow \infty$ we find
\begin{equation}
G \approx -F \sim \left\{ \begin{array}{ll}
\exp \{ -[gx^2/2 + mx - (2g)^{-1}\varepsilon^2\ln x] \},\;\;& \alpha
=1 \; ,\\
\exp \{ -[g(\alpha+1)^{-1} x^{\alpha +1}+mx] \},\;\;& \alpha >1 \; .
\end{array}
\right.
\end{equation}

The first formula (with $g=1 $ and $m=0$) explains the structure of
asymptotic expansion (30).

\section{Symmetry properties of the spectrum of the Dirac equation}

\hspace{.5cm}
Here we discuss symmetries of solutions of the Dirac equation for
particles of zero mass.

It is clear that  when both $m$ and $U$ are zero the Dirac
equation (and the corresponding term in the Lagrangian) is chirally
symmetric, i.e., does not change under a global transformation:
\begin{eqnarray}
\Psi \rightarrow \exp(i\alpha \gamma_5) \Psi,~~~
\bar{\Psi} \rightarrow \bar{\Psi} \exp(i \alpha \gamma_5) \;.
\end{eqnarray}

{}From the point of view of the spectrum the chiral symetry means that all
states are parity degenerate; i.e., the masses of the states $0^+$ and
$0^-$ (or $1^+$ and $1^-$) are  the same.

 It is easy to demonstrate that system (25) with $m=0$ is invariant
under even more general transformations when $U\not=  0$:
\begin{eqnarray}\nonumber
E\rightarrow E,~~ \kappa \rightarrow -\kappa,~~ U\rightarrow -U,~~
 V \rightarrow V,\\ G(r)\rightarrow -F(r),~~ F(r) \rightarrow G(r).
\end{eqnarray}

  It follows from (42) that for $U\equiv 0$   the spectrum is degenerate
with respect to the sign of the Dirac quantum number $\kappa$; i.e., it
depends on only the total momentum $j$, not on the way $l$ and $s$ are
coupled (chiral degeneracy).

Another symmetry of the Dirac equation is

\begin{eqnarray}
E \rightarrow -E,~~ \kappa \rightarrow -\kappa,~~ U\rightarrow U,~~ V
\rightarrow -V,~~ G(r)\leftrightarrow  F(r),
\end{eqnarray}
which in contrast to (42) connects  states with positive and negative
energy. In particular, for $V\equiv0$ (scalar interaction) there is a
doubling of states with a given $\mid E_n \mid, E_n=\pm \mid E_n \mid$, and
we can always consider $E_n>0$.  In the specific case $E=0$ (zero modes)
there is, at  first sight, a chiral degeneracy of states. However,
it is easy to show that the degenerate states with $E=0$ do not exist at
all.  Multiplying the first equation in (25) by $G$,
multiplying the second  by $F$,
and integrating the difference from $r=0$ to $r=\infty$, we obtain an
identity:
\begin{eqnarray}
\int\limits ^{\infty}_{0} (G^2+F^2) \frac{dr}{r} =
2\kappa^{-1}\int\limits^{\infty}_{0} (E-V) GFdr
\end{eqnarray}
(we have used in the derivation the circumstanse that $G$ and $F$ vanish
at both  $r=0$ and $r=\infty$).  From Eq.$\;$(44) one can see that for the
purely scalar interaction there are no  solutions with zero
energy.

Note that the symmetry transformation, Eq.$\;$ (42), can be also presented
in an operator form. For a zero-mass particle the Hamiltonian $H$ and the
Dirac operator $K$ are \begin{equation} H=\mbox{\boldmath
$\alpha$}{\bf p} + \beta U(r)+V(r),\;\; K=-\beta(\mbox{\boldmath$\sigma$}
{\bf l}+1) \end{equation} (the quantum
number $\kappa$ is the eigenvalue of the operator $K$). It can be easily
seen that in this case we have \begin{equation} \gamma_5 H\gamma_5 =
\mbox{\boldmath$\alpha$}
{\bf p} - \beta U(r)+V(r)\;\;, \;\; \gamma_5 K\gamma_5=-K,
\end{equation}
which are the same as Eqs. (42).
%\end{document}

\section{Results of numarical calculations}

\hspace{.5cm}
Here we report results of our numerical calculations. In Fig.4 is shown
 the dependence of the eigenvalues $\varepsilon_{n \kappa} = E_{n
 \kappa}/M$ on the ratio $\zeta/ \mid \kappa\mid$ for  potential
(27) ( the solid lines correspond to the lowest states for a given value of
 $\kappa$, $n=1;$ the dashed lines are for the first excited states,
$n=2$).  The energy eigenvalues decrease monotonically with  growing
Coulomb parameter $\zeta$, and for $\zeta \rightarrow \mid  \kappa
\mid$ they have a square-root singularity. The latter is
characteristic of potentials with a Coulombic behavior at $r
\rightarrow 0$ and is connected to the phenomenon of the collapse to the
 center [10,11].

In Fig. 4 one can see that the levels with $ \kappa>0$ lie much
 higher than those with $ \kappa<0$ (for a given $n$). The physical
 meaning of this effect becomes clear when one recalls that the
 centrifugal barrier is proportional to $ \kappa(\kappa+1)$ and
 is absent for $ \kappa=-1$ (for example), in contrast to the case of $
\kappa=+1$.

Note that the energies of the  lowest $(n=1)$ states with $\kappa<0$ reach
the value $\varepsilon =0$ at the maximal possible value of the Coulomb
coupling constant $\zeta = -\kappa \;$ (Fig. 5). All the other states
also have  square-root singularities at $\zeta =|\kappa|$, but
their energies are positive (Fig. 6). The numerical values of
$\varepsilon_{n\kappa} (\zeta)$ in the two extreme cases $\zeta=0$
and $\zeta = |\kappa|$ are given in Table I, where the dependence
of $\varepsilon_{n\kappa}$ on the quark mass is also illustrated. Figure 7
shows the energies of several lowest levels within the interval
$0.3<\zeta<0.8$, which is
of practical interest for the $u\bar{b}$ system \footnote{As can be seen
from Figs. 5 and 6, the dependence of the energy $\varepsilon_{n\kappa}$ on the
light-quark mass is practically linear. This is in agreement with ref.
 [27], where the energy spectrum of Eq.$\;$ (25) was calculated at $\zeta
 =0.6$ with hyperfine splitting taken into account, which is necessary for a
 detailed comparison with the experimental spectrum (the procedure was
 similar to the calculation of the hyperfine structure for relativistic
 Coulomb problem).  The value of hyperfine splitting obtained in ref. [27]
 differs from experimental data for charmonium and bottomonium by over
 30\%.}.

 We note that  the singularity at $\zeta = | \kappa|$
can be removed when one introduces a cutoff of the Coulomb potential $V(r)
= - \zeta/r$ at small distances\footnote{In the case of the lowest levels
with $ \kappa<0$, for which $\varepsilon ( \kappa) \rightarrow 0$ in the
limit $\zeta \rightarrow \mid \kappa\mid$ (see Fig. 4), this singularity can
be derived analytically --- see Eq.$\;$ (50) below.}:

\begin{equation}
V(r) = \left\{ \begin{array}{ll}
-\zeta r^{-1},              ~~ & \mbox{if}~~~  r>r_0,\\
(-\zeta/r_0) f(r/r_0),~~ & \mbox{if}~~~  0<r<r_0,
\end{array}
\right.
\end{equation}
where $r_0$ is the cutoff radius, and $f(0)<\infty$.

 In the case  $r_0 \ll M^{-1}$ the influence of the
 cutoff on the energy levels is important only in the immediate
 vicinity of the point $\zeta = \mid \kappa\mid$. Here the level sinks to the
boundary $\varepsilon =0$ (it corresponds to a boundary $\varepsilon =-m$
for $m \not=0$) at some value $\zeta =\zeta_{n \kappa} >\mid \kappa\mid$,
which depends both on $r_0$ and on the concrete form of the cutoff function
$f(r/r_0)$ in Eq.$\;$(47).

With a further increase of $\zeta$ the level goes on down to the
region of negative values of  energy, but the spectrum stays discrete
(this is in contrast to the Coulombic problem with a vector potential
[10,11]).  Thus pair creation does not take place.

\section{Exact solution for $E=0$}

\hspace{.5cm}
In QED the solutions of the Dirac equation simplify considerably for $E=\pm m$,
which corresponds in our case $(m=0)$ to $E=0$. Eqs.$\;$ (28) in terms of
linear combinations $u=G+F,~ U=G-F$ assume the form ($\varepsilon =0$)

\begin{equation}
u'-xu +\frac{\kappa+\zeta}{x} v=0 , \mbox{~~~}
v'+xv+\frac{\kappa-\zeta}{x} u=0 .
\end{equation}

At $\zeta =-\kappa$ the first equation can be solved explicitly:
$$u= C \exp (x^2/2),$$ which yields $C=0$ and hence $G=-F=v/2 $.
The normalized wave functions are\footnote{This result can be easily
generalized
to the case $m \not= 0$ and arbitrary scalar potential $U(r)$.}

\begin{equation}
G(x)= -F(x) =\pi^{-1/4} \exp (-x^2/2) .
\end{equation}

Using the perturbation theory in the parameter
$(\kappa^2-\zeta^2)^{1/2}\ll 1$, one can determine the  behavior of
the energy near $\zeta = - \kappa$:
\begin{equation}
\varepsilon(\zeta) = \pi^{-1/2}(1-\zeta^2/\kappa^2)^{1/2}+...
\end{equation}
(for details see Appendix A).
Computations confirm this asymptotic expansion; see Fig. 5.

Note that the solution of the type in (49) exists only for the states with
$\kappa <0$. In the case  $\zeta=\kappa$, Eqs.$\;$ (48)
have a nonzero solution $$G=F= \mbox{const} \cdot \exp (x^2/2)$$ which is not
admissible because  of the exponential growth at infinity.

These results can be generalized to arbitrary values of $\zeta$ and
$\kappa$.  Solving  (48) for one of functions $u,v$, we come to the
equation
\begin{equation}
w^{\prime \prime}+\frac{1}{x} w' -\left[\pm 2+x^2
+\frac{\kappa^2-\zeta^2}{x^2}\right]w=0 ,
\end{equation}
 where the upper
(lower) sign corresponds to the function $u\;(v)$. We  consider the case
$\zeta > \mid \kappa \mid$, when the cutoff of the Coulomb part
is essential. A solution of these equations for $r>r_0$, decreasing at
infinity, can be expressed in terms of Whittaker functions,
\begin{equation}
u=C_1x^{-1} W_{-1/2,ig}(x^2)  ,\mbox{~~~}
v= C_2x^{-1} W_{1/2, ig}(x^2) ,
\end{equation}
where $g =(\zeta^2-\kappa^2)^{1/2} >0$, and $C_1,C_2$ are
constants.

 Insertion of (52) into the first of Eqs.$\;$ (48) yields the connection
between $C_1,C_2$:  \begin{equation} C_1= 2(\kappa+\zeta)C_2 .\end{equation}

In the internal region, $0<r<r_0$, the Dirac equation should be solved with
cutoff potential (47), where one can neglect the linear potential because
of the relation $r_0\ll1/M$.  For the simplest case, $f(x)\equiv 1$ (a
square-well cutoff), the solution can be found analytically and can be
expressed in terms of Bessel functions with half-integer index.

The energy spectrum is found from the joining of the internal and external
solutions at $r=r_0$. In the case of the Dirac equation one can join the
ratio $F/G$ instead of the logarithmic derivative.

As a result we have
\begin{equation}
2(\kappa+\zeta) W_{-1/2,ig}(x^2_0)/W_{1/2,ig}(x_0^2) =\xi ,
\end{equation}
where $\xi = u(x_0)/v (x_0)$ is determined from the interval solution and
depends on $ \kappa,\zeta$ and the cutoff model. For the states with
$\kappa=\mp 1$ in the square-well cutoff case, we have
\begin{equation}
\xi \mid
_{\kappa=-1}=-\frac{1}{\xi}\Bigl|_{\kappa=1}=\frac{1-\zeta(1+\mbox{\rm
ctg}\zeta)}{1+\zeta(1-\mbox{\rm ctg} \zeta)} .  \end{equation} Equation (54)
can
be solved numerically; it determines the dependence of the Coulomb constant
$\zeta_{n\kappa}$, which corresponds to the descent of the $n$-th level to the
boundary $\varepsilon = 0$, on the cutoff radius $r_0$ [for a certain choice of
the cutoff function $f(r/r_0)$].

 Finally, we note that a solution (49) can be generalized to the case of
nonzero quark  mass  and an arbitrary scalar potential $U(r)$. There is an
exact solution of the Dirac equation  with $\zeta =
-\kappa $ and $E=0$,
\begin{equation}
G_0(r) = - F_0(r) = N \exp\left[-\int\limits^r_0 m(r')dr'\right]\;,
\end{equation}
where $m(r) = m+U(r) $ is a variable quark mass and $N$ is
the normalization constant, [see Eq.$\;$(A.8)]. Since this wave
function has no nodes, this solution corresponds to the
lowest $(n=1)$ level with the fixed $\kappa$.  The
energy of the state, as well as other physical
quantities, has a square-root singularity at
$\zeta\rightarrow -\kappa$, characteristic of the
relativistic Coulomb problem:
\begin{equation}
E(\zeta) = c_1\beta + O(\beta^2)\;, \;\; \rho =
1-c_2\beta +...,
\end{equation}
where $\beta = (1-\zeta^2/\kappa^2)^{1/2}$, and $\rho$ is
the relative weight of the lower component of the Dirac
bispinor,
\begin{equation}
\rho = \int\limits^{\infty}_0 F^2(r) dr
\Bigl/\int\limits^{\infty}_0G^2(r)dr .
\end{equation}
It can be shown that
\begin{equation}
c_1=N^2\;, \;\; c_2=4N^2\langle r\rangle
\end{equation}
(see Appendix A), where $\langle r\rangle $ is the mean radius of the
bound state (56). The effective-potential method
 given in Appendix A is most useful in deriving these formulas, as well as
for a quantitative analysis of the wave functions near $\zeta = \pm \kappa$.

As is seen from Eq.$\;$ (57), at $\zeta\approx |\kappa|$ the
parameter $\rho$ is close to one (see also Table II). This means that the
motion of a light quark is definitely relativistic. The
coefficients $c_1$ and $c_2$ for a particular case of
$m(r) =m+\sigma r$ are presented in Fig. 8.
Finally, let us note that square-root singularities of the type of
(57) are directly connected with the behavior of the Coulomb potential at
$r\rightarrow 0$ and disappear when it is regularized at small distances.

\section{Discussion and conclusion}

\hspace{.5cm}
In conclusion we discuss the structure of the resulting spectrum,
briefly compare it to experiment, and summarize our results\footnote{The
main results of this paper  were presented in ref. [8].}.

We start with the nonrelativistic spectrum and use the formulas given in
Appendix B, which yield the spectrum shown in Fig. $9,a$. Here
the splitting between the $S = 1$ and $S = 0\; (L = 0)$ levels is due to the
hyperfine interaction and, as is seen in Appendix B, is proportional to
$(m_1m_2)^{-1}$. When both masses are large $(m_i \gg M = \sigma^{1/2})$,
all splittings are small, including the spin--spin [$O ((m_1m_2)^{-1}$)] and
spin--orbit splittings, which contains terms $O(m^{-2}_1)$, $O(m^{-2}_2)$ and
$O((m_1m_2)^{-1})$, which one can denote as $(LS)_1$, $(LS)_2$ and
$(LS)_{12}$ --- see Appendix B.

In the case  $m_2 \gg m_1 \sim M \;$ (Fig. $9,b$), the intervals of
$(LS)_1$ become the largest, and the spectrum transforms in such a way that
the coupling of the spin of particle 2 becomes very weak. In the limit
$m_2 \rightarrow \infty$ it finally decouples, leading to the Isgur--Wise
symmetry [5]. Namely, the states can be classified by the
total momentum of the light particle $j$, while the states of total
momentum of the system ${\bf J} = {\bf j} + {\bf S}_2$ are degenerate with
respect to the direction of ${\bf S}_2$.

This is what one observes in the Dirac spectrum (see Fig. 4 of our
numerical calculations and Fig. $9,b$, which give a schematic
description of the Dirac spectrum). Here the degeneracy (Isgur--Wise
symmetry) is complete.

This should be compared to the experimental picture in Fig. $9,c$, where for
the charmed mesons the order of the levels is the same as in Fig. $9,b$ ,
 but the splittings are still large. For $B$-mesons the splitting of
lowest levels is smaller by a factor of 3 $\;$ (52 MeV), as it should be,
since the mass of the $c$ quark is $\approx 3$ times smaller than that of
$b$ quark.  We also note that the experimental splitting $\Delta E\sim 450$
MeV between $j=1/2$ and $j=3/2$ states is reproducible in
our results when one takes $\sqrt{\sigma}\sim 0.5$ GeV and $\zeta \sim
0.6\div 0.8$.

Thus we can conclude that the Dirac equation yields a reasonable
qualitative description of the actual spectrum. We hope to discuss
this point in more detail in future publications, where we will also give
predictions for the $D_S$-, $B$-, and $B_S$-mesons.

Summarizing our results,
we have derived the Dirac equation from the Feynman--Schwinger representation
of
the quark--antiquark Green's function in case of the color Coulomb interaction.

Assuming that the confining interaction
 can be introduced into the Dirac equation in the sameway as to the
Coulomb interaction, we have clarified the Lorentz structure of the
confining interaction connecting a light quark to a heavy antiquark.

The analysis of the solutions of the Dirac equation shows that a
potential growing at infinity  yields confinement only if it is
a scalar,  not the time component of a 4-vector. In this aspect there
is an essential difference from the nonrelativistic case.

We have also studied the dependence of the charge $\zeta$ in the critical
region, $\zeta \sim |\kappa |$, and found its dependence on
the cutoff of the Coulomb potential. We have computed energy eigenvalues
for several lowest levels and have compared them with the nonrelativistic
spectrum and experimental results.

After this paper had been finished and submitted for publication,
we learned of several papers [28--30] in which analogous problems are
discussed\footnote{The authors are indebted to  A.E.Kudryavtsev who
pointed out to us the above mentioned papers.}.

In the paper by Ono [29], a relativistic generalization of the Richardson
potential was considered. The Coulomb part of the potential was
considered as a 4-vector component, and its confining part  as a
relativistic scalar. The paper by Dremin and Leonidov [30] discusses the
properties of solutions of the Dirac equation with  scalar and vector
coupling. Using the squared Dirac equation, the authors arrived at the
conclusion that it is only in the case of a scalar potential growing at
infinity that bound states can exist in the system. That conclusion is in
full agreement with the results of our Sec.$\;$ 4, where we study in greater
detail the asymptotics of the Dirac wave functions at infinity and give a
physical explanation of the above results using the effective potential
method.

%\newpage
\noindent
\hfill {\it APPENDIX A}
\centerline{\bf Method of effective potential}\\

\setcounter{equation}{0}
\def\theequation{A.\arabic{equation}}

The system of Eqs.$\;$ (25) can be reduced to an equivalent Schr\"{o}dinger
equation with a potential depending on energy and having, in general, a
rather complicated form \footnote{See refs. [10,24,26] on
physical phenomena near the boundary of the lower continuum $E=-mc^2$
in QED [10], where the above method was successfully applied.}.
 There is a  considerable   simplification if $\zeta =|\kappa| = 1,2,...\;.$
 This case we shall discuss further.

 First let $\zeta=-\kappa$. Proceeding in Eqs.$\;$ (27) to the linear
 combinations $u = G+F,\;\; v=G+F,$ we obtain equations from which
 the function $v$ is easily excluded. Then we get

\begin{eqnarray}
\nonumber
u^{\prime \prime}+[\varepsilon^2-1-(x+\mu)^2+2\zeta\varepsilon
x^{-1}]u&=&0,\;\;u(0)=0,\\
\int\limits^{\infty}_0\left(1+\frac{\zeta}{\varepsilon
x}\right)u^2(x)dx& =&\frac{1}{4},
\end{eqnarray}
i.e., the Schr\"{o}dinger equation with
the effective energy $\tilde{E}$ and potential $\tilde{U}$
\begin{equation}
\tilde{E}=\frac{1}{2}(\varepsilon^2-1),\;\;
\tilde{U}=\frac{1}{2}(x+\mu)^2-\varepsilon\zeta x^{-1} \;.
\end{equation}
 Note that the boundary condition at
zero for $u(x)$ follows from the fact that we have $A/B = -1$ in Eq.$\;$
(29), and the normalization condition in (A.1) corresponds to the usual
condition $$\int\limits_0^{\infty} (G^2+F^2) dr =1 $$ for the discrete
spectrum.

In particular, for $\varepsilon =0$ we have $ \tilde{E}=-1/2$, which cannot
be an eigenvalue for the harmonic oscillator. Therefore, we have $u(x)
\equiv 0$ and $v'+(x+\mu)v=0$, from which the exact  solution (49) follows.

For the case $\zeta = \kappa$  the function $v$, but not $u$,
satisfies an equation of the like (A.1), and we have
$\tilde{E}=(\varepsilon^2+1)/2$.  Here we have $\tilde{E} =1/2$  with
$\varepsilon=0$; therefore we have $v(x)\equiv 0$ (with the boundary condition
at zero taken into account), while all nonzero solutions $u(x)$ increase
exponentially at infinity. Thus, the Dirac equation has no physically
acceptable solutions with zero energy if $\zeta =\kappa >0$.

The following results are obtained from (A.2). Let $E_n=E_n(\alpha, \mu)$ be
the energy spectrum of the Schr\"{o}dinger equation with the potential
\begin{equation}
V(r)= \frac{1}{2}(r+\mu)^2 -\alpha r^{-1}\;,
\end{equation}
 with $0<r<\infty$ and $l=0$
[in particular, $E_n(0,0)=2(n-1/4),\;\; n=1,2,3,...]\;$. Then the eigenvalues
$\varepsilon_n^{(\mp)}$ of the problem considered may be obtained from the
transcendental equation:
\begin{equation}
E_n(|\kappa|\varepsilon, \mu)=\frac{1}{2}(\varepsilon^2 \pm \mbox {sgn}~\kappa)
\end{equation}
(the $\pm$ sign coincides with the sign of
the quantum number $\kappa$). Due to the relations
 $$\frac{\partial E_n}{\partial \alpha} = -\left\langle
\frac{1}{r}\right\rangle
 \; ,~~ \frac{\partial E_n}{\partial \mu} =\langle r\rangle +\mu \; ,$$
 sometimes called the Hellmann--Feynman theorem, the energies $E_n(\alpha,\mu)$
decrease monotonously as the Coulomb parameter $\alpha$ goes up and increase
with the growing $\mu$. Taking this into account, we get \begin{equation}
\varepsilon_1^{(-)}=0<\varepsilon_1^{(+)}<\varepsilon_2^{(-)}<\varepsilon_2^{(+)}<...
\end{equation}
 (with fixed $|\kappa|$). The results of the numerical solution are
given in Table I.

The results are easily generalized to the case of an arbitrary scalar
potential $U(r)$. Instead of (A.1) we get
\begin{equation}
\frac{d^2\chi }{dr^2}+\left\{E^2-[m+U(r)]^2+(\mbox {sgn}~\kappa)
\frac{dU}{dr}+\frac{2|\kappa|E}{r}\right\}\chi =0\;,
\end{equation}
 where $\chi(0)=0$, and the normalization condition does not
depend explicitly on the form of $U(r)$:
\begin{equation}
\int\limits^{\infty}_0\left(1+\frac{|\kappa|}{Er}\right)\chi^2(r)dr=\frac{1}{4}
\end{equation}
 $[\chi=u(r)$ for the states with $\zeta=-\kappa$ and
$\chi=v(r)$  for those with $\zeta =\kappa$].

At $\zeta=-\kappa$ there is an exact solution of the Dirac equation which
corresponds to zero energy [and does not depend explicitly on $\kappa$;
see Eqs.$\;$ (54)]; here $\chi_0(r)\equiv 0$,
\begin{eqnarray}
\nonumber
G_0(r)=-F_0(r) = N \exp \left[-\int\limits^r_0 m(r')dr'\right],\\
N=\left\{ 2\int\limits^{\infty}_0 dr\exp \left[-2 \int\limits^r_0
m(r')dr'\right]\right\}^{-1/2}.  \end{eqnarray}

The behavior of the energy of the state at $\zeta\rightarrow -\kappa$ can
be also determined. Assuming
\begin{eqnarray}
E=c_1\beta +O(\beta^2),~~~\beta=(1-\zeta^2/\kappa^2)^{1/2}\rightarrow 0,\\
\nonumber
u= \beta u_1(r)+...\;,~~~ v = v_0+\beta v_1(r)+...\;,
\end{eqnarray}
 and substituting these  expansions into (25), we obtain
\begin{eqnarray}
\frac{du_1}{dr}-mu_1+c_1v_0 =0,\;\; \frac{dv_1}{dr}+mv_1 =
\frac{2|\kappa|}{r} u_1.
\end{eqnarray}

 The solution of the first equation which decreases at infinity has the
 form
\begin{equation}
u_1(r) = 2c_1N\exp \left[\int\limits^r_0 m(r')dr'\right] \int\limits^{\infty}_r
dx
\exp \left[-2\int\limits^x_0 m(x')dx'\right], \end{equation}
  while the function $v_1(r) $ is calculated by a quadrature; at $r=0$ we have
 $u_1(0)=c_1/ N,~ v_0(0)=2N$. Let $$\xi(r) \equiv u/v=\beta
 u_1/v_0+O(\beta^2);$$ then it follows from Eq.$\;$ (28) that $$\xi(0)
 =\frac{A+B}{A-B}
 =\frac{(1+\beta)^{1/2}-(1-\beta)^{1/2}}{(1+\beta)^{1/2}+(1-\beta){1/2}}=
 \frac{1}{2} \beta +O(\beta^2)$$
 whence $u_1/v_0|_{r=0}=1/2$, and finally we get Eq.$\;$ (59) for $c_1$.

 Thus for the states whose energy vanishes at $\zeta=-\kappa$, the
 coefficient of the square-root singularity in $E(\zeta)$ can be found
 explicitly  for an arbitrary scalar potential $U(r)$. Other physical
 quantities, such as $\rho$, have the same singularities at
 $\zeta=-\kappa$.

 Using Eqs.$\;$ (25) and (A.6), we find for the parameter $\rho$ introduced
 in (58) that \begin{eqnarray} \nonumber
 \rho =\frac{1-s}{1+s}\;,~~ s=2\int\limits^{\infty}_0
uvdr\Bigl/\int\limits^{\infty}_0 (u^2+v^2)dr= \\
%\nonumber
=\int\limits^{\infty}_0 m(r)\chi^2 dr\Bigl/\int\limits^{\infty}_0
 \left(E+\frac{|\kappa|}{r}\right)\chi^2 dr=
 \frac{4}{E}\int\limits^{\infty}_0m(r) \chi^2 dr
\end{eqnarray}
[here $\zeta=|\kappa|$, and in
 the latter formula $\chi(r)$ is normalized according to Eq.$\;$ (A.7)].
 Hence, expansion (57) for $\rho$ follows, where $$c_2=2\int\limits^{\infty}_0
 u_1v_0 dr.$$ Substituting in the above expressions for $u_1, v_0$, and
 changing the order of integration, we obtain
 \begin{equation}
 c_2=8c_1^2\int\limits^{\infty}_0 \exp
 \left[-2\int\limits^r_0m(r')dr'\right]rdr, \end{equation} which coincides with
  Eq.$\;$ (59).

 Let us consider a few simple dependences $m(r)$.

 1) A constant mass $m(r) = m$ corresponds to QED. At $\zeta=-\kappa$, we
 have
  \begin{equation}
  G_0=-F_0=m^{1/2}e^{-mr},~~~~~c_1=m,~~~~~c_2=2~,
 \end{equation}
  which corresponds to the Sommerfeld formula for the relativistic Coulomb
 problem, which has the following form for the $\kappa =-n$ states
 ($j=l+1/2=n-1/2$):
\begin{equation}
E_{nj}(\zeta) = m\left(1-\frac{\zeta^2}{n^2}\right)^{1/2}\equiv m\beta,~~~~
\rho= \frac{1-\beta}{1+\beta}.
\end{equation}

 The levels with radial quantum numbers $n_r>0$ also have the Coulomb
 singularity at $\zeta \rightarrow |\kappa|$, but the corresponding energies
 are positive. For example, for the $ns_{1/2 }$ states $(k=-1, n_r=n-1$):
 \begin{equation}
 E_{n,1/2} = m \left(\frac{n-1}{N}+\frac{\beta}{N^3}+...\right),~~ \rho_n =
 (N+n-1)^{-2}\left(1-\frac{2n}{N}\beta+...\right) ,
  \end{equation}
  where $\beta = (1-\zeta^2)^{1/2},~~ \zeta = Z\alpha\rightarrow 1$, and
 $N=(n^2-2n+2)^{1/2}$.

  At $n\gg1$, we find $\rho_n=1/4 n^2 \rightarrow 0,$ and the bound states
  become nonrelativistic.

 2) At $m(r) = m+\sigma r$, we find
\begin{equation}
 G_0=-F_0=N \exp \{-(\sigma r^2/2+ mr)\},
\end{equation}
\begin{eqnarray}
 \nonumber
 N=(\sigma/\pi)^{1/4}[\exp(\mu^2)\cdot \mbox{erf}~c(\mu)]^{-1/2},
 ~~\mu=m/\sigma^{1/2}, \\ c_1=N^2,~~~c_2=4\sigma^{-1}c_1(c_1-m)
 \end{eqnarray} (note that the dimensionless coefficient $c_2$ depends only on
  $\mu$). In the extreme case at $\mu\ll1$ we obtain
 \begin{equation}
  c_1=\left(\frac{\sigma}{\pi}\right)^{1/2}\left(1+\frac{2\mu}{\pi^{1/2}}+...
\right),~~
   c_2=\frac{4}{\pi}\left(1+\frac{4-\pi}{\pi^{1/2}}\mu+...\right),
 \end{equation} while at $\mu\gg 1$ we obtain \begin{equation}
c_1=m(1+\frac{1}{2}\mu^{-2}+...),~~ c_2=2-\mu^{-2}+...  \end{equation}
Numerical
results for $c_1,c_2$ are given in Fig. 9.

%\newpage
\noindent
\hfill{\it APPENDIX B}\\

\setcounter{equation}{0}
\def\theequation{B.\arabic{equation}}

The nonrelativistic spectrum of two particles with all spin interactions
taken into account can be obtained in the perturbative way ($1/m$ expansion):
%\nonumber
$$m=m_1+m_2+E(n)+\Bigl\langle n \mid
\left(\frac{\mbox{\boldmath$\sigma$}_1 {\bf L}}{m_1^2}+ \frac{\mbox{\boldmath
$\sigma$}_2{\bf L}}{m_2^2}\right) \left(\frac{1}{4r} \frac{d\varepsilon}{dr}
+\frac{1}{2r}\frac{dV_1}{dr}\right)+$$%\\ \nonumber\\
$$+\frac{(\mbox{\boldmath$\sigma$}_1+\mbox{\boldmath
$\sigma$}_2){\bf L}}{2m_1m_2}~ \frac{1}{r}
\frac{dV_2}{dr}+ \frac{\mbox{\boldmath$\sigma$}_1
\mbox{\boldmath$\sigma$}_2}{12m_1m_2} V_4(r)+$$%\\

%\nonumber\\
%\begin{eqnarray}
\begin{equation}
+\frac{1}{12m_1m_2} (3 \mbox{\boldmath$\sigma$}_1
{\bf n} \cdot \mbox{\boldmath$\sigma$}_2{\bf n}
- \mbox{\boldmath $\sigma$}_1 \mbox{\boldmath$\sigma $}_2)V_3(r)
\mid n\Bigr\rangle.
\end{equation}
%\end{eqnarray}

The contributions of the scalar interaction $U(r) =\sigma r$
and the vector interaction $V(r) = -4\alpha_s/3 r$  to
 the spin splittings are\\
 $$\frac{1}{r} \frac{d\varepsilon}{dr}
  =\frac{\sigma}{r} + \frac{4}{3} \frac{\alpha_s}{r^3}\;,~~~
  \frac{1}{r} V'_1 = - \frac{\sigma}{r}\;,~~~
  \frac{1}{r} V'_2 = \frac{4}{3} \frac{\alpha_s}{r^3}\;,$$\\
  \begin{equation}
  V_4 = \frac{32\pi \alpha_s}{3} \delta^{(3)}({\bf r})\;, \mbox{~~~}
  V_3 = \frac{4\alpha_s}{r^3} .
  \end{equation}

  The main difference appears in $V'_1$, where only the scalar
  interaction enters. Note that the overall signs of the spin--orbit terms
  due to the scalar
   $(\sigma r)$ and
    vector $(-4\alpha_s/3r)$ potentials are different.
%\end{document}
 \newpage

\begin{center}
TABLE I.
The eigenvalues $\varepsilon_{n\kappa} $ of the Dirac equation with $\zeta=
0$ and $\zeta=|\kappa|$\\

\vspace{1.0cm}

\begin{tabular}{|l|r|l|l|r|r|} \hline
\multicolumn{2}{|c|}{ } &
\multicolumn{2}{c|}{ $\zeta =0$} &
\multicolumn{2}{c|}{ $\zeta =|\kappa|$} \\ \hline
$n$&$\kappa$ &$\mu=0$ &$\mu=0.3$ &$\mu=0$&$\mu=0.3$\\ \hline
1&-1&1.61944&1.84441& .00000& .00000\\
2&-1&2.60263&2.80689&1.06901&1.17491\\
3& -1& 3.29118& 3.49080& 1.96846& 2.11712\\
4& -1& 3.85541& 4.05300& 2.67856& 2.84277\\
 & & & & & \\
 1&-2& 2.14652&2.36761& .00000&.00000\\
 2&-2&2.95197& 3.15853&.64443&.70949\\
 3&-2 &3.57353& 3.77508&1.39768&1.51077\\
 4& -2&4.09947&4.29854&2.08531&2.22342\\
 & & & & &\\
 1&-3&2.56927&2.78850&.00000&.00000\\
 2&-3&3.26852&3.47647&.45001&.49567\\
 & & & & &\\
 1&-4&2.93218&3.15029&.00000&.00000\\
 & & & & &\\
 1& 1&2.29403 &2.49206&.64015&.79011\\
 2&1&3.03103&3.22747&1.62588&1.79736\\
 3&1&3.62598&3.82161&2.39019&2.57021\\
 & & & & &\\
 1&2&2.70440&2.90645&.36916&.46149\\
 2&2&3.35376&3.55217&1.12003&1.25059\\
 & & & & &\\
 1&3&3.05967&3.26183& .25449&.31931\\
 & & & & &\\
 1&4&3.40866& 3.60000& .19328& .24284\\ \hline
 \end{tabular}
\end{center}

\newpage

\begin{center}
TABLE II.
Values of the parameter $\rho$ in Eq.$\;$ (58)

\vspace{1.0cm}
\begin{tabular}{|l|r|l|l|r|r|} \hline
\multicolumn{2}{|c|}{ } &
\multicolumn{2}{c|}{ $\zeta =0$} &
\multicolumn{2}{c|}{ $\zeta =|\kappa|$} \\ \hline
$n$&$\kappa$ &$\mu=0$ &$\mu=0.3$ &$\mu=0$&$\mu=0.3$\\ \hline
1&-1&.15534&.13156& 1.00000& 1.00000\\
2&-1&.20341&.17727&.50983&.44930\\
3& -1& .21280& .18976& .35668& .31909\\
4& -1& .21640& .19580& .30697& .27815\\
 & & & & & \\
 1&-2& .16165&.14188& 1.00000&1.00000\\
 2&-2&.19556& .17401&.66958&.61953\\
 3&-2 &.20662& .18651&.47148&.43447\\
 4& -2&.21178&.19318&.38478&.35496\\
 & & & & &\\
 1&-3&.16444&.14718&1.00000&1.00000\\
 2&-4&.16601&.15049&1.00000&1.00000\\
 & & & & &\\
 1&1&.22266&.18839&.34861&.31937\\
 & & & & &\\
 1&2&.20986 &.18360&.54374&.51578\\
 & & & & &\\
 1&3&.20203&.18024&.65694&.63317\\
 & & & & &\\
 1&4&.20532&.17791&.72679&.70663\\
 \hline
\end{tabular}
\end{center}

\newpage
\centerline{\bf Figure captions.}

Fig. 1. Quark--antiquark Green's function describing transition from
the initial state $\Gamma (y,\bar{y})$ to the final state
$\Gamma (x,\bar{x})$. Wavy lines refer to the quark (antiquark)
Green's function $\Delta_1(x,y)$~($\Delta_2(x,y)$)

Fig. 2. Trajectory corresponding to virtual pair creation.
$a$ --- Scalar case; $b$ --- vector case

Fig. 3. Nonrelativistic potential $V_{NR}(r)$ ($a$) and effective potential
$U_{eff}(r)$ ($b$) for various Lorentz structures of the interaction.
Curves {\it 1--3} correspond to the three choices in Sec. 4

Fig. 4. Energy spectrum for the Dirac equation with potential (27). The
solid curves correspond to the $n=1$ states, the dashed ones to $n=2$.
The curves are labeled with the values of $\kappa$

Fig. 5. The eigenvalues $\varepsilon_{1\kappa}(\zeta),~ n=1$, of the Dirac
equation with    potential (27) in the vicinity of $\zeta=-\kappa$.
The solid curves correspond to the $\kappa=-1$ states, the
dotted ones to $\kappa=-2.$ The curves are labeled with the values of
$\mu=m/M$

Fig. 6. The same as in the preceding figure, for excited states
($\kappa=-1, n=2$ and $\kappa=1, n=2$)

Fig. 7. The energies of the lowest levels, $\varepsilon_{n\kappa}$, versus
the Coulomb parameter $\zeta$ (for a quark mass $m=0$)

Fig. 8. Dependence of $c_1$  and $c_2$ [Eq.$\;$ (59)] on the  quark mass

Fig. 9. Energy spectrum for the heavy quark--light antiquark system
(qualitative).  $a$ --- $ m_1, m_2 \gg M=\sigma^{1/2}$;
$b$ --- $m_1\rightarrow 0, m_2\rightarrow \infty$ ;  $c$ --- experiment.
The energy scale in part $a$ is much larger then in parts $b$ and $c$

\end{document}